\documentclass[traditabstract]{aa} 
\usepackage[authoryear]{natbib}
\usepackage{graphicx}
\usepackage{amsmath}
\usepackage{amssymb}
\usepackage{txfonts}
\begin{document}
\title{Giant outburst of EXO 2030+375: pulse-phase resolved analysis of
       \textsl{INTEGRAL} data}


\author{D. Klochkov\inst{1}\and
  A. Santangelo\inst{1}\and
  R. Staubert\inst{1}\and
  C. Ferrigno\inst{2}
  }

\institute{
  Institut f\"ur Astronomie und Astrophysik, Universit\"at T\"ubingen (IAAT), 
  Sand 1, 72076 T\"ubingen, Germany
  \and
  INAF IFC-Pa, via U. La Malfa 153, 90146 Palermo, Italy
  }

\date{Received ***; accepted ***}

 
\abstract{
In June--September 2006 the Be/X-ray binary EXO~2030+375 experienced the
second giant outburst since its discovery. The source was shown to
have a complicated pulse-averaged X-ray spectral continuum with possible
evidence of cyclotron absorption features. In this paper we present 
the first pulse-phase resolved analysis of the broad band X-ray spectra of 
EXO~2030+375 obtained with the \textsl{INTEGRAL} observatory close 
to the maximum and during the decay phase of the giant outburst. 
We report a strong variability of the spectrum with pulse phase. 
Alternative spectral continuum models are discussed. The dependence of the 
spectral parameters on pulse phase during the maximum of the outburst 
and the evolution of the pulse profiles with time 
are qualitatively consistent with the 
pulsar's emission diagram changing from the fan-beam geometry close to 
the maximum of the outburst to a combination of pencil and fan beams 
(of comparable intesities) at the end of the decay phase. 
Evidence of a cyclotron absorption line around 63\,keV at the pulse phase 
interval preceeding the main peak of the pulse profile is present in 
the spectrum obtained close to the maximum of the outburst. 
}


\maketitle
%
\section{Introduction\label{intro}}

The transient accreting pulsar EXO 2030+375 belongs to the most common
type of X-ray pulsar systems -- the Be/X-ray binaries.
Such systems form a subclass of high mass X-ray binaries. 
They consist of a pulsar and a Be (or Oe) companion, 
a main-sequence star of spectral type B (or O) that shows Balmer 
emission lines \citep[see e.g.][for a review]{Slettebak88}.
The line emission is believed to be associated with an equatorial outflow 
of material expelled from the rapidly rotating Be star that probably 
forms a quasi-Keplerian disk around its equator 
\citep{Hanuschik96,Quirrenbach_etal97}. If the disk reaches a radius 
comparable to the periastron separation, then disk gas accreted by the 
neutron star can power a significant (and usually transient) X-ray source.

Be/X-ray binaries typically show two types of outburst behavior:
\begin{enumerate}
\item Normal (or type I) outbursts.
They are characterized by relatively low X-ray luminosities
$L_{\rm X} \sim 10^{36}{\rm-}10^{37}\,{\rm erg~s}^{-1}$, low spin-up rates
(if any), and recurrence at the orbital period (or its multiples). Such
outbursts last from days to weeks and are associated with the periastron
passages of the neutron star.
\item Giant (or type II) outbursts.
They are characterized by higher X-ray luminosities $L_{\rm X} \gtrsim
10^{37}\,{\rm erg~s}^{-1}$ and high spin-up rates. Such outbursts occur
irregularly. They last several weeks and are not correlated with any
particular orbital phase. The typical time between outbursts 
is around
several years. Giant outbursts are thought to stem from a dramatic 
expansion of the disk surrounding the Be star, leading to the formation 
of an accretion disk around the compact object.
\end{enumerate}

EXO 2030+375 is one of the best-studied Be/X-ray binaries. It was 
discovered with the \textsl{EXOSAT} satellite during a giant outburst
in 1985 \citep{Parmar_etal89a}. The optical companion of the pulsar is 
a B0 Ve star identified by optical and infrared observations of the 
\textsl{EXOSAT} error circle \citep{MotchJanot87,Janot_etal88,Coe_etal88}.
The orbital period and eccentricity of the system are $\sim$46\,d and
$\sim$0.42, respectively \citep{Wilson_etal02}. The period of X-ray 
pulsations is $\sim$42\,s. There were two giant outbursts in the history 
of observations of EXO~2030+375. During the first one in 1985 (when the 
pulsar was discovered), the X-ray luminosity of the source reached a value of
$L_{\rm 1-20~keV}\sim 2 \times 10^{38}\,\text{erg}\,\text{s}^{-1}$
(assuming a distance of 7.1 kpc, \citealt{Wilson_etal02}). The
spin frequency of the pulsar changed dramatically, with
a spin-up timescale $-P/\dot P \sim 30$\,yr indicating the
formation of an accretion disk around the neutron star. 
The second giant outburst took place in June--September 2006 
\citep{CorbetLevine06,Klochkov_etal07} and was again accompanied by 
a strong spin-up of the neutron star. The X-ray luminosity at the 
maximum of the outburst was slightly lower than during the 1985 
giant outburst: 
$L_{\rm 1-20~keV}\sim 1.2 \times 10^{38}\,\text{erg}\,\text{s}^{-1}$
\citep{WilsonFinger06}. 

During the 2006 giant outburst the source
was observed several times with the \textsl{INTEGRAL} satellite.
A preliminary pulse-averaged spectral analysis of some of
these observations are presented in \citet{Klochkov_etal07}. 
A detailed analysis of pulse-averaged \textsl{RXTE} spectra obtained during
the outburst was performed by \citet{Wilson_etal08}. It has been shown
that the pulse averaged X-ray continuum of the source has a complicated
shape and cannot be modeled by a simple power law/cutoff model.  
\citet{Wilson_etal08} include an absorption line at $\sim$10\,keV 
(which they interpreted as a cyclotron line) in their spectral model, whereas
\citet{Klochkov_etal07} have shown that the spectrum can be fitted 
equally well without the absorption line, but including a broad emission 
``bump'' at $\sim$15\,keV. 

Here we present for the first time pulse-phase resolved
broad band (3--150\,keV) spectra of EXO~2030+375 during a giant outburst.
For our analysis we used all the available \textsl{INTEGRAL} data taken during
the 2006 outburst. The X-ray continuum of the source shows strong variability 
with pulse phase, with some features present only at particular 
pulse-phase intervals. 
The description of observations that we used is provided 
in Sect.~\ref{obs}. Details of data processing are described in 
Sect.~\ref{data}. Sections~\ref{max} and \ref{decay} are devoted to the 
analysis of the data close to the maximum and during the decay of the 
outburst, respectively. The results are discussed in Sect.~\ref{discuss} 
and briefly summarized in Sect.~\ref{concl}.

\section{Observations\label{obs}}

\begin{figure}
\centering
\resizebox{\hsize}{!}{\includegraphics{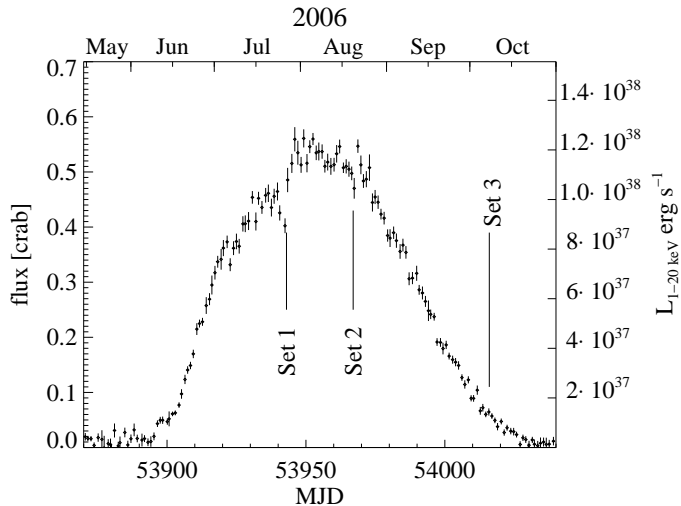}}   
\caption{The \textsl{RXTE/ASM} light curve of EXO~2030+375 showing the
  2006 giant outburst. Times of \textsl{INTEGRAL} observations analyzed in
  this work are marked with vertical lines. The observations are referred
  to as Set~1, Set~2, and Set~3 in the text.}
\label{exoobs}
\end{figure}

The \textbf{INTE}rnational \textbf{G}amma \textbf{R}ay \textbf{A}strophysics
\textbf{L}aboratory (\textsl{INTEGRAL}, \citealt{Winkler_etal03})
performed three pointed observations of EXO~2030+375 during its
giant outburst in June--September 2006. Two observations 
(on 19--20 August and 23--25 September) were done close to the maximum 
of the outburst. The third one (on 6--8 October) was performed
at the end of the decay phase, when the X-ray luminosity
dropped by a factor of $\sim$10 with respect to the maximum of the
outburst. A part of the \textsl{RXTE/ASM} light curve\footnote{We used
the results provided by the ASM/RXTE team.} including the giant outburst
is shown in Fig.~\ref{exoobs}. The times of the \textsl{INTEGRAL} observations
are indicated and referred to as Set~1, Set~2, and Set~3 throughout 
the paper. 

Observations corresponding to Sets 1 and 2 were done
when the source was at similar luminosity levels
($L_{\rm 1-20~keV}\sim 10^{38}\,\text{erg}\,\text{s}^{-1}$) before 
and after the maximum of the outburst. A preliminary analysis of the 
two data sets was presented by \citet{Klochkov_etal07} who showed
that spectral and timing characteristics of the source during the two
observations were similar. Due to the high X-ray flux, the data allowed us
to perform a detailed pulse-phase resolved spectral analysis of EXO~2030+375
using twelve phase bins. During the observations corresponding
to Set~3, the X-ray luminosity of the source was at a level typical of
normal (Type I) outbursts by this system 
($L_{\rm 1-20~keV}\sim 10^{37}\,\text{erg}\,\text{s}^{-1}$).
Spectra and pulse profiles of the source during these observations
were considerably different with respect to those in Sets~1 and 2.
In spite of a poorer statistics, the data taken in Set~3 still 
allowed us to perform pulse-phase resolved spectral analysis, although with a 
coarser binning (four bins) in pulse phase. Table~\ref{obslist} 
contains the summary of the observations analyzed in this work.

\begin{table}
\caption{Summary of observations.}
\label{obslist}
\centering
\begin{tabular}{c c c c}
\hline\hline
Obs.  & Obs.time  & Mean & Mean ASM        \\
      &  (ks)     &  MJD & flux (mCrab)    \\
\hline
& & &  \\
Set 1 & 62        & 53942.9 & 450 \\
Set 2 & 140       & 53967.6 & 500 \\
Set 3 & 122       & 54015.2 & 60 \\
\hline
\end{tabular}
\end{table}

\section{Data processing\label{data}}

For our analysis we used the data obtained with the instruments 
\textsl{IBIS/ISGRI} (20--300\,keV, \citealt{Ubertini_etal03}) and 
\textsl{JEM-X} (3--30\,keV, \citealt{Lund_etal03}) onboard
\textsl{INTEGRAL}. To perform the standard data reduction, the Off-line
Science Analysis (OSA) software (version 6) was used
\citep{Courvoisier_etal03}. To construct energy-resolved pulse profiles we 
used the software developed at IASF, Palermo \citep{Mineo_etal06}.
While extracting spectra and pulse profiles of EXO~2030+375, 
two other bright sources in the field of view of the \textsl{INTEGRAL} 
instruments, Cyg~X-1 and Cyg~X-3, were included to the 
extraction catalog used by the analysis software,
which removes the contamination of EXO~2030+375. All other sources in
the field of view are much weaker than EXO~2030+375 and therefore do
not provide any noticeable contamination.
The spectral analysis of the observations was performed using the
\texttt{XSPEC} v.11.3.2l spectral fitting package \citep{Arnaud96}.
Following the OSA User 
Manuals\footnote{\texttt{http://isdc.unige.ch/index.cgi?Support+documents}},
we added a systematic error to the final count rates at a level of 1\% for 
\textsl{ISGRI} and 2\% for \hbox{\textsl{JEM-X}} to account for 
small-scale uncertainties in the response matrices of the respective 
instruments. 

\begin{table}
\caption{The pulse periods and associated derivatives for three sets of 
observations analyzed in this work. }
\label{tabper}
\centering
\begin{tabular}{c c c c}
\hline\hline
Observation & Reference   & $P$ & $-dP/dt$        \\
            & epoch (MJD) & [s] & [$10^{-8}$ s/s] \\
\hline
& & &  \\
Set 1 & 53966.392169 & $41.519270(5)$ & $2.758(20)$  \\
Set 2 & 53942.240864 & $41.58084(3)$  & $2.96(21)$   \\
Set 3 & 54014.010200 & $41.45425(6)$  & $0.22(23)$   \\
\hline
& & & \\
\multicolumn{4}{c}{Note: uncertainties in parentheses (68\%) 
refer to the last digit(s).}
\end{tabular}
\end{table}

To extract X-ray pulse profiles and to define pulse phase 
intervals for pulse phase resolved spectra, all photon arrival 
times were translated into the
solar system barycenter and corrected for orbital motion in the
binary. The orbital parameters were taken from \citet{Wilson_etal05}:
$P_{\rm orb}=46.0202(2)$\,d, $T_{\rm peri}={\rm MJD}\,51099.43(2)$, 
$a\sin i=238(2)$\,lt-sec, $e=0.416(1)$, $\omega =210^{\circ}.8(4)$.
The pulse periods and associated derivatives were determined
for each of the three observational sets individually by employing initial 
epoch-folding and a subsequent phase-connection analysis similar to 
\citet{Ferrigno_etal06} and \citet{Deeter_etal81}. To perform phase-connection,
we used a pattern-matching technique applied to well-defined pulse profiles 
from a sufficient number of pulses. Any variation in the pulse shape 
inside each observation is marginal and does not affect our method. 
The determined periods and period derivatives are summarized in 
Table~\ref{tabper}. Values corresponding to Sets~1 and 2 are slightly
improved with respect to those reported in \citet{Klochkov_etal07}.

\section{Maximum of the outburst\label{max}}

\begin{figure*}
  \centering
  \includegraphics[width=17cm]{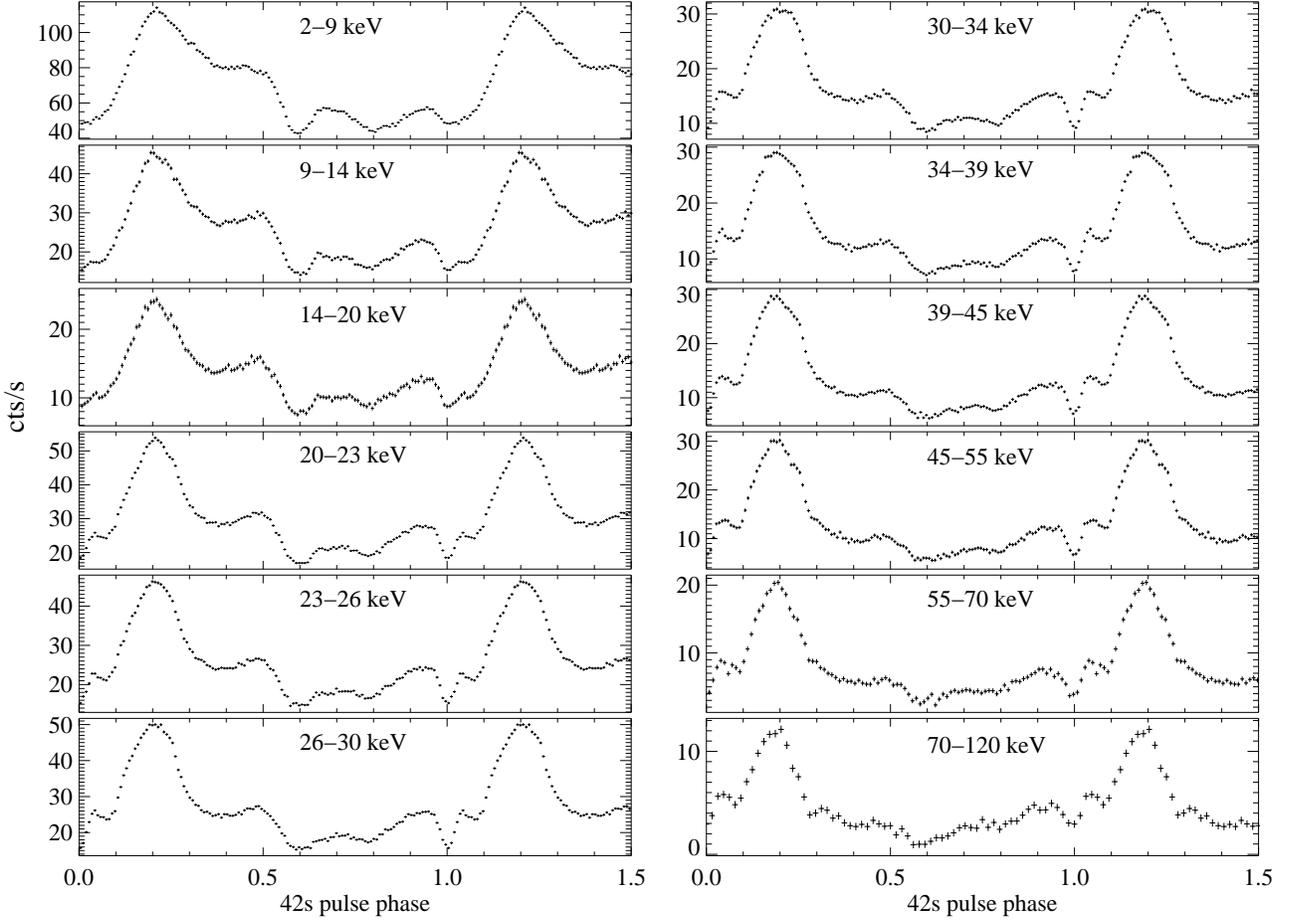}
  \caption{Energy-resolved pulse profiles of EXO~2030+375 obtained with
    \textsl{JEM-X} (below 20\,keV) and \textsl{IBIS/ISGRI} (above 20\,keV)
    in Set 2 (where the data have the best statistics).}
  \label{ppenres}
\end{figure*}

Observations corresponding to Sets 1 and 2 were performed close to the maximum
of the outburst (Fig.~\ref{exoobs}). During Set~1, which is roughly half 
as long as Set 2 (see Table~\ref{obslist}), the main target of
observations was Cyg~X-3 located $4^\circ$ away from EXO~2030+375. 
Therefore, the statistics of the EXO~2030+375 data obtained during Set~1
is noticeably worse than during Set~2. As a result, our pulse-phase 
resolved analysis of the maximum of the outburst is mainly driven by the data 
taken in Set~2.

To explore spectral changes with pulse phase, we constructed
pulse-phase resolved spectra and energy-resolved pulse profiles. While
the former allow detailed study of spectral parameters as a function of 
pulse phase, the latter can be used to explore the intensity of different 
pulse-profile components (peaks, dips, etc.) in different energy ranges 
without relying on any particular spectral function.

\subsection{Energy-resolved pulse profiles}

\begin{figure}
\centering
\resizebox{\hsize}{!}{\includegraphics{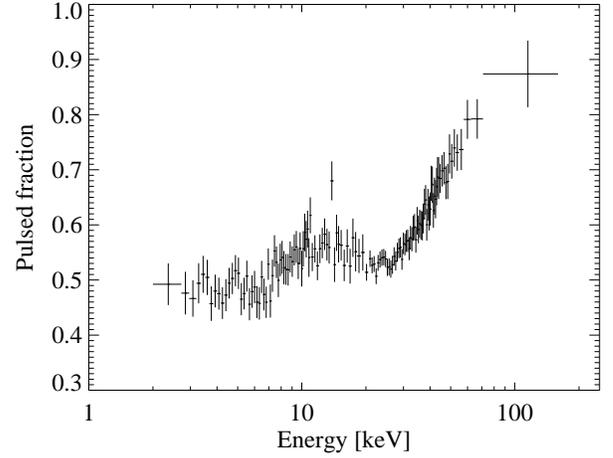}}   
\caption{The pulse fraction of EXO~2030+375 as a function of energy}
\label{pfrac}
\end{figure}

Figure~\ref{ppenres} presents energy-resolved background-subtracted 
pulse profiles of the source obtained during the observations corresponding 
to Set~2. The corresponding pulse period and its derivative, as well as the 
zero epoch, are provided in Table~\ref{tabper}. The pulse profiles in
the energy bands below 20\,keV were obtained from the \textsl{JEM-X} 
data, while the \textsl{IBIS/ISGRI} data were used for energies 
above 20\,keV. Several components can be distinguished in the profiles: 
the main 
peak at phase $\sim$0.2 followed by a fainter component at phase $\sim$0.5 
(which we will refer to as the ``trailing shoulder'') and two smaller peaks 
in the interpulse interval 0.6--1.0. The sharpest detail of the profile 
is a narrow dip preceding the main peak. 
Phase 0.0 was arbitrarily chosen to coincide with the dip
in the average \textsl{ISGRI} pulse profile. 
As one can see, the shape of the profile 
changes smoothly with energy indicating variations in the X-ray spectrum with
pulse phase. 
The pulse fraction of the source determined as 
$(F_{\rm max}-F_{\rm min})/(F_{\rm max}+F_{\rm min})$ (where $F_{\rm max}$
and $F_{\rm min}$ are fluxes in the maximum and minimum of the pulse
profile, respectively) as a function of energy is shown in Fig.~\ref{pfrac}. 
One can see a bump-like structure around $\sim$10--20\,keV that is probably
related to the continuum feature in this energy 
range (the ``bump'' or the absorption line, see Sects.~\ref{intro} 
and \ref{max_prs}). Above $\sim$20\,keV, the pulse
fraction seeply increases.

\subsection{Pulse-resolved spectra\label{max_prs}}

\begin{figure}
\centering
\resizebox{\hsize}{!}{\includegraphics{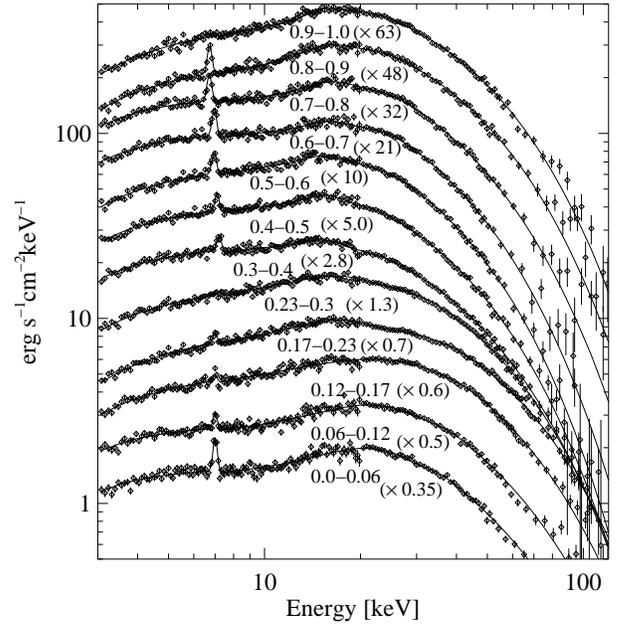}}   
\caption{Pulse-resolved broad band X-ray spectra of EXO~2030+375 obtained in 
  Set~2. Phase bins are indicated (phase zero is the same as in 
  Fig.~\ref{ppenres}). Values in parenthesis indicate the 
    multiplicative factor applied to the flux in each 
    spectrum to avoid overlaps. The solid line represents 
  the fit of the spectra using Model I (see text for the model details).}
\label{speprs}
\end{figure}

As one can conclude from the dependence of the pulse profile 
shape on energy, the X-ray spectrum of EXO~2030+375 clearly varies with
pulse phase. We have therefore performed a separate analysis of 
the spectra accumulated in different pulse phase intervals.
Phase binning was chosen to provide similar statistics of spectra
in each bin and to have better phase resolution around the main peak where
the most rapid spectral changes are expected. 
The binning was chosen {\em a priori} based solely 
on the shape of the pulse profile. No further adjustments were made 
after the spectra had been obtained.
Figure~\ref{speprs} shows unfolded pulse phase-resolved spectra of the source
(shifted vertically with respect to each other to avoid 
overlaps). The pulse phase is defined in the same way as in the previous 
section. Variability of the spectral continuum is clearly seen.

As pointed out in Sect.~\ref{intro}, the pulse-averaged spectral 
continuum of the source is rather complicated and cannot be modeled by 
any of the simple spectral functions (a power law modified 
at higher energies by an exponential cutoff), which are usually used to 
fit spectra of accreting pulsars. In an attempt to model the spectrum
\citet{Wilson_etal08} modified a power law/cutoff model by a
Gaussian absorption line at $\sim$10\,keV, while \citet{Klochkov_etal07}
included a broad Gaussian emission component at $\sim$15\,keV instead.
In this work we tried to use both models to fit the pulse-resolved
spectra. The reduced $\chi^2$ in all phase bins is very similar for the 
two models. This does not allow one
to ultimately choose between the two spectral functions. The two models are 
referred to as Model I (with a ``bump'') and Model II (with an absorption 
line at 10\,keV) throughout the paper.

Below we describe the spectral functions corresponding to the two models.
The common part of Models~I and II is the power law/cutoff continuum
smoothed at the cutoff energy by a third-order polynomial:
\begin{equation}
I_{\rm cont} = K\cdot
\begin{cases}
E^{-\Gamma}, & \text{if\,} E \leq E_{\rm cutoff}-\Delta E \\
E^{-\Gamma} \cdot \exp{\left(-\frac{E-E_{\rm cutoff}}{E_{\rm fold}}\right)}, 
          & \text{if\,} E > E_{\rm cutoff}+\Delta E \\ 
AE^3 + BE^2 + CE + D, 
          & \text{if\,} E_{\rm cutoff}-\Delta E < E < \\
          & < E_{\rm cutoff}+\Delta E,
\end{cases}
\label{cont}
\end{equation}
where $K$ is the normalization coefficient, $E$ the photon energy; 
$\Gamma$, $E_{\rm cutoff}$, $E_{\rm fold}$, and $\Delta E$ are model 
parameters. Numerical coefficients $A$, $B$, $C$, and $D$ are chosen 
to obey the condition of continuity for the function and its first 
derivative at the points $(E_{\rm cutoff}-\Delta E)$ and 
$(E_{\rm cutoff}+\Delta E)$. Model~I includes a broad Gaussian 
emission component (a ``bump'') around 15\,keV:
\begin{equation}
I_{\rm Model\,I}=I_{\rm cont}+ K_{\rm bump}\exp\left\{-\frac{(E-E_{\rm bump})^2}{2\sigma_{\rm bump}^2}\right\},
\label{ModelI}
\end{equation}
where $E_{\rm bump}$ and $\sigma_{\rm bump}$ are the energy and width of the
``bump'', and $K_{\rm bump}$ is the numerical constant describing the
intensity of the component. In Model~II, the power law/cutoff continuum
$I_{\rm cont}$ is modified by a multiplicative absorption line with a Gaussian
optical depth profile:
\begin{equation}
I_{\rm Model\,II}=I_{\rm cont}\cdot \exp\left\{-\tau_{\rm line}
\exp\left(-\frac{(E-E_{\rm line})^2}{2\sigma_{\rm line}^2}\right)\right\},
\label{ModelII}
\end{equation}
where $E_{\rm line}$, $\sigma_{\rm line}$, and $\tau_{\rm line}$ are the 
centroid energy, width, and the central depth of the line, respectively.
Additionally, we multiplied the functions $I_{\rm Model\,I}$ and 
$I_{\rm Model\,II}$ by the factor $\exp[-N_{\rm H}\sigma_{\rm bf}(E)]$ 
describing the low-energy absorption by cold matter in the line of sight.
Here, $\sigma_{\rm bf}(E)$ is the photoabsorption 
cross-section per hydrogen atom 
for matter of cosmic abundances \citep{Balucinska92} used in the
\texttt{phabs} model of \texttt{XSPEC}, and $N_{\rm H}$ is the equivalent 
hydrogen column density. We also added a Gaussian emission line 
to model the iron fluorescence line at $\sim$6.4\,keV. The latter, however,
improves the fit only slightly. For all pulse-phase resolved spectra the
value of $N_{\rm H}$ was fixed to $2\times 10^{22}\,{\rm cm}^{-2}$ (the
average of the best-fit values found in all phase bins). It was also
found that the energy of the ``bump'' in Model I does not change
significantly with pulse phase being close to 15\,keV. It was, therefore,
fixed to this value for all phase bins. 

\begin{figure}
\centering
\resizebox{\hsize}{!}{\includegraphics{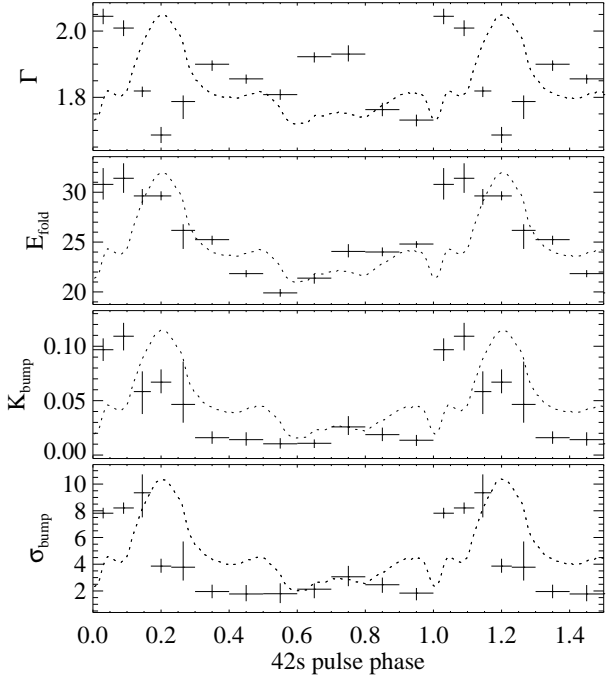}}   
\caption{Best-fit parameters of Model\,I (see text) as a function 
    of pulse phase obtained by fitting the spectra from Set 2.}
  The dotted curve shows the averaged \textsl{ISGRI} (20--120\,keV) pulse 
  profile scaled and shifted vertically to match the plot. 
  Vertical error bars correspond to 90\%-uncertainties.
\label{parprs}
\end{figure}

\begin{figure}
\centering
\resizebox{\hsize}{!}{\includegraphics{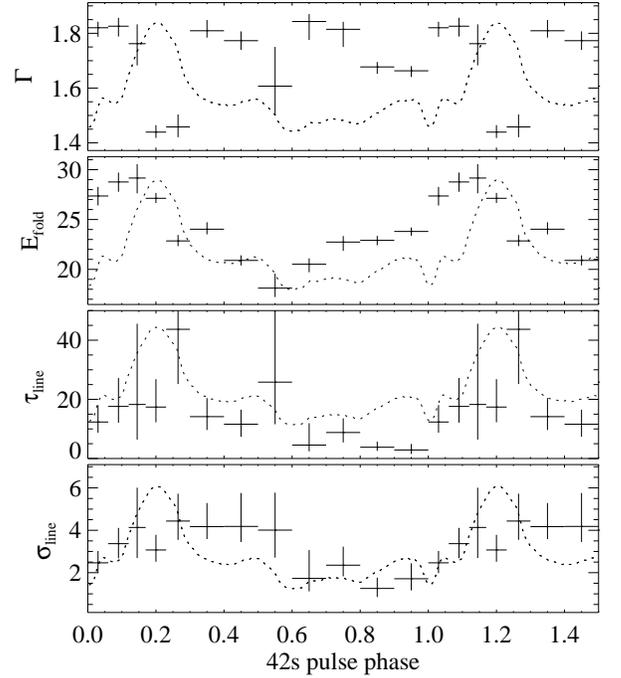}}   
\caption{Best-fit parameters of Model\,II (see text) as a function 
    of pulse phase obtained by fitting the spectra from Set 2. As in 
    Fig.~\ref{parprs}, the dotted curve shows the \textsl{ISGRI} pulse 
    profile. Vertical error bars correspond to 90\%-uncertainties.}
\label{parprs_wf}
\end{figure}

Figures~\ref{parprs} and \ref{parprs_wf} show the most important spectral 
parameters obtained using the two models as a function of pulse
phase. The data correspond to the observations performed in Set~2, which
have the best statistics. 
Vertical error bars correspond to 90\%-uncertainties.
Note that the pulse phase range shown in the figures is from 0.0 to 
1.5 so that some data points are repeated (the same is true for 
Fig.~\ref{parprs_too2}).

The same kind of analysis was performed using the observations 
corresponding to Set~1 (slightly before the maximum of the outburst). 
Due to poorer statistics, we had to use a coarser binning compared to Set~2. 
The spectral parameters of the pulse-resolved spectra were 
less constrained in this case. However, the behavior of the parameters 
agrees with what is found in Set~2. We also notice that the best-fit 
parameters of the pulse-averaged spectra taken in Sets~1 and 2 are also 
consistent with each other (see Table~3 in \citealt{Klochkov_etal07}).

\begin{figure}
  \centering
  \resizebox{\hsize}{!}{\includegraphics{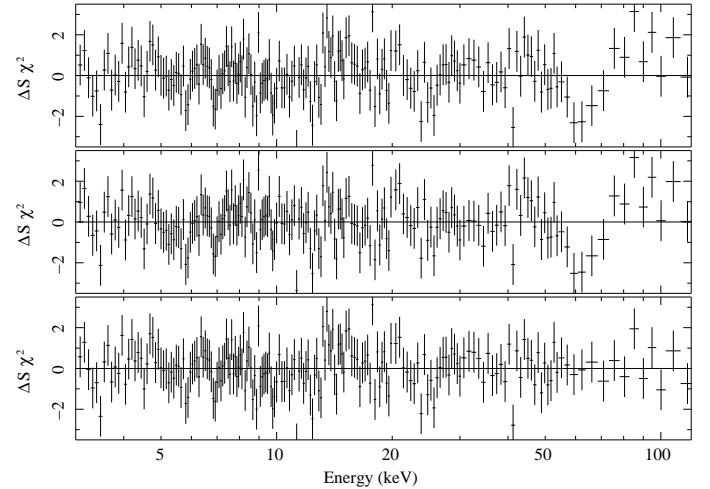}}   
  \caption{Residuals of the spectrum from the phase bin 0.06--0.12 
    fitted with Model I (top) and Model II (middle). The bottom 
    panel shows the residuals after fitting the spectrum with Model I
    where a Gaussian absorption line at $\sim$63.6\,keV is included
    (the corresponding fit with Model II looks very similar).}
  \label{residuals}
\end{figure}

Residuals left after fitting the spectrum by Models~I and II from 
phase bin 0.06--0.12 obtained in Set~2 reveal an absorption feature at
$\sim$63\,keV (see Fig.~\ref{residuals}). Inclusion of a Gaussian
absorption line at this energy flattens the residuals. The best-fit
energy of the line is $63.6_{-2.9}^{+3.7}$\,keV for Model I and
$63.3_{-2.9}^{+4.0}$\,keV for Model II. We checked the presence 
and the energy of the feature using other possible continuum models, such as 
Fermi-Dirac cutoff \citep{Tanaka86} and the so-called negative and positive
power-law times EXponential model (NPEX, \citealt{Makishima_etal99}),
including a ``bump'' or an absorption line at 10\,keV to match the continuum.
It was found that the presence and the energy of the feature are 
independent of the choice of the spectral function. The F-test
probability that the line is due to statistical fluctuations
is $\sim$3$\times 10^{-5}$ \citep[see however][about 
non-applicability of the F-test to line-like features]{Protassov_etal02}. 

\section{Decay of the outburst\label{decay}}

\subsection{Energy-resolved pulse profiles}

\begin{figure}
  \centering
  \resizebox{\hsize}{!}{\includegraphics{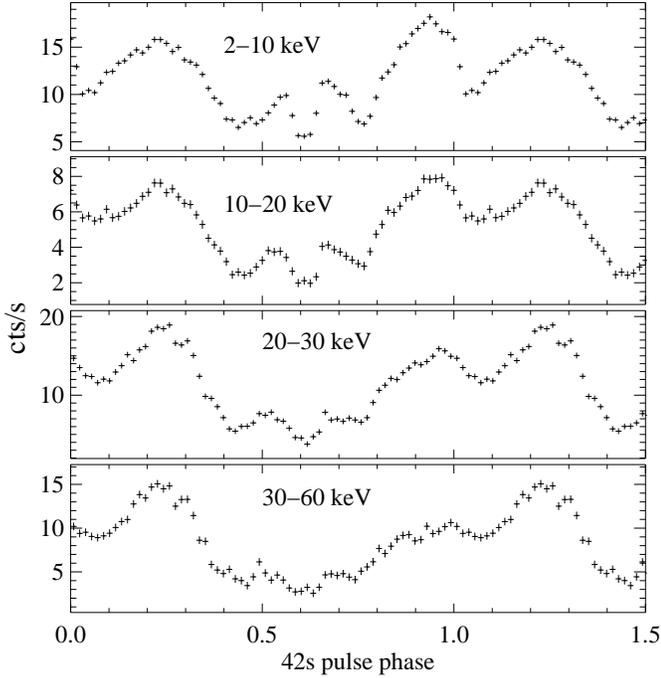}}   
  \caption{Energy-resolved pulse profiles of EXO~2030+375 obtained with
    \textsl{JEM-X} (below 20\,keV) and \textsl{IBIS/ISGRI} (above 20\,keV)
    in Set~3 (at the end of the outburst).}
  \label{ppenres_too2}
\end{figure}

As mentioned in Sect.~\ref{obs}, during Set~3, \textsl{INTEGRAL}
caught the source at the end of the outburst's decay phase when the X-ray 
luminosity was $\sim$10 times lower than the maximum of the
outburst (Fig~\ref{exoobs}). Due to lower statistics we used
much coarser binning in energy (for pulse profiles) and pulse phase 
(for pulse-phase resolved spectra) than in Set~2
(4 instead of 12 in both cases).
The resulting pulse profiles are shown in Fig.~\ref{ppenres_too2}.
As before, the corresponding pulse period, its derivative, and the 
epoch zero can be found in Table~\ref{tabper}. To choose 
the time of phase zero in accordance with the one used for Set~2 we used
the dependence of the pulse profiles shape on the X-ray luminosity
presented in Fig.~1 of \citet{Parmar_etal89b}. The dependence allows one
to identify features in the profiles (e. g. the dip around
pulse phase 0.6) in both observations (Sets~1 and 2)
and choose the zero epoch for both observations consistently. 

The dependence of the pulse profile on energy in Set~3 
(Fig.~\ref{ppenres_too2}) is mainly characterized
by a decreasing relative amplitude of the peak at phase $\sim$0.9
with respect to the main peak (phase $\sim$0.2) with energy.
The sharp dip seen in the 2--10\,keV at phase $\sim$0.6 almost disappears
at higher energies. Generally, one can see that the shape of the pulse 
profiles is substantially different as compared to the profiles obtained
close to the maximum of the outburst at similar energies. 
In Sect.~\ref{discuss} we show that the changes could be qualitatively
explained if assuming that a pencil beam component appears at the end of the
outburst.

\subsection{Pulse-phase resolved spectra}

\begin{figure}
  \centering
  \resizebox{\hsize}{!}{\includegraphics{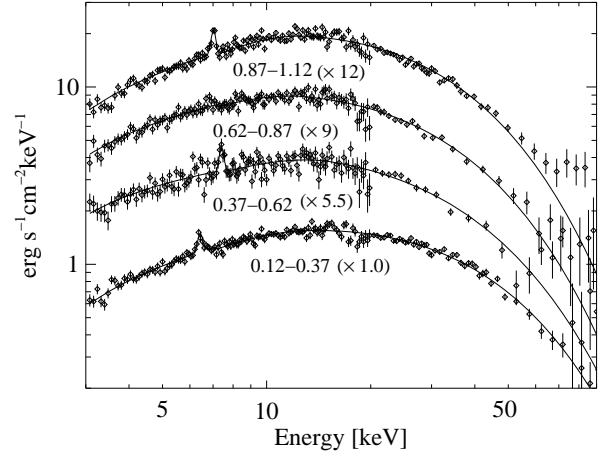}}   
  \caption{Pulse-phase resolved broad band X-ray spectra of EXO~2030+375 
    obtained in Set~3. Phase bins are indicated (phase zero is the same as in 
    Figs.~\ref{ppenres_too2}). Values in parentheses indicate the 
    multiplicative factor applied to the flux in each 
    spectrum to avoid overlaps. The solid line represents 
    the fit of the spectra using the power law/cutoff model (see text for the 
    model details).}
  \label{speprs_too2}
\end{figure}

To produce pulse-phase resolved spectra from the data obtained in Set~3, 
we used four phase bins. They cover each of the two peaks and two phase 
intervals of the interpulse. The spectral continuum can be modeled well 
by the power law/cutoff function provided by Eq.~\ref{cont}
without inclusion of the ``bump'' or the 10\,keV absorption line. 
Like in Sect.~\ref{max_prs}, we modified the model at lower energies by 
photoabsorption (with $N_{\rm H}\sim 2\times 10^{22}\,{\rm cm}^{-2}$) and 
added a Gaussian line to model the iron fluorescence emission at 
$\sim$6.4\,keV.
Figure~\ref{speprs_too2} represents the unfolded pulse-phase resolved spectra
fitted with the described model and shifted vertically with respect to each 
other in order to avoid overlaps. (This was done multiplying the flux
in each spectrum by a numerical factor indicated in parentheses.) 
Significant variation in the continuum
is clearly seen. The pulse phase is defined in the same way as in 
Fig.~\ref{ppenres_too2}.

\begin{figure}
  \centering
  \resizebox{\hsize}{!}{\includegraphics{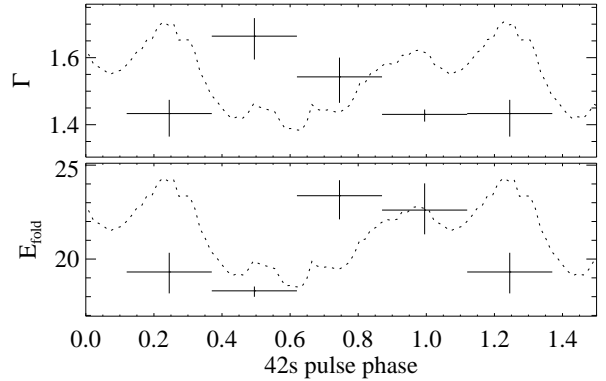}}   
  \caption{Best-fit spectral parameters of EXO~2030+375 obtained in Set~3 
    as a function of pulse phase. 
    The dotted curve shows the averaged \textsl{ISGRI} (20--120\,keV) pulse 
    profile. Vertical error bars correspond to 90\%-uncertainties.}
  \label{parprs_too2}
\end{figure}

Figure~\ref{parprs_too2} shows the photon index $\Gamma$ and the folding
energy $E_{\rm fold}$ as a function of pulse phase during Set~3.
Vertical error bars correspond to 90\%-uncertainties. 

\section{Interpretation and discussion\label{discuss}}

\subsection{Pulse-period derivatives}

The values of pulse-period derivatives measured close to the maximum
and at the end of the outburst (Table~\ref{tabper}) basically
confirm the statement in \citet{Klochkov_etal07} that the spin-up
rate of the pulsar is proportional to its X-ray flux as expected
in the simple accretion torque theory \citep[e. g.][]{PringleRees72}.
The observed dependence is similar to what was measured by \textsl{EXOSAT}
during the 1985 giant outburst of EXO~2030+375 \citep{Parmar_etal89a}.
However, with only a few points, it is not possible to explore the 
correlation using more complicated accretion torque models. 

\subsection{Variation in the spectral continuum with pulse 
phase\label{discvar}}

The \textsl{INTEGRAL} observations have shown that the spectral 
continuum of EXO~2030+375 is strongly pulse-phase dependent. Observations 
of the source close to the maximum of the outburst allowed us to perform 
a detailed study of this dependence. A harder main peak and softer 
interpulse region observed in the source is a common property of
accreting pulsars \citep[see e.g.][end references therein]{Tsygankov_etal07}.
It is usually explained by assuming that, during the peak, we mostly
see the comptonized photons coming from a hot region close to the 
footstep of the accretion column while in the interpulse
softer radiation scattered by the upper parts of the column is observed
\citep{BaskoSunyaev76}. A closer look at Fig.~\ref{ppenres}, however, reveals 
a more complicated picture. Both the main peak and its left
flank increase in relative amplitude with energy,
leading to a slight shift of the profile maximum towards an earlier
pulse phase. This shows that the spectral continuum changes asymmetrically 
with respect to the main peak. The variation in spectral parameters
with pulse phase (Figs.~\ref{parprs} and \ref{parprs_wf}) 
demonstrates the corresponding complex pulse-phase dependence of the spectrum.
The photon index $\Gamma$ reaches a minimum (the hardest spectrum) in the 
middle of the main peak (top panel in Figs.~\ref{parprs} and \ref{parprs_wf}), 
while the maximum of the exponential folding energy $E_{\rm fold}$ is shifted
with respect to the peak towards earlier pulse phases. This contrasts
to e.g. \hbox{Her~X-1}, where both the maximum of $E_{\rm fold}$ and the
minimum of $\Gamma$ are coincident with the main pulse
\citep{Klochkov_etal08}. 

A possible qualitative explanation for the observed asymmetry of the 
spectral variation with pulse phase in EXO~2030+375 can be the following. 
Since the luminosity of the source close to the maximum of the 
outburst, $\sim$10$^{38}\,{\rm erg\,s}^{-1}$, is well above the local 
Eddington limit (so-called critical luminosity, 
$L_c \sim 10^{37}\,{\rm erg\,s}^{-1}$, 
\citealt{BaskoSunyaev76,Staubert_etal07}), it is expected that accreted 
matter is decelerated by radiation pressure leading to the formation
of an optically thick accretion column. The bulk of the emission in
this case is expected to occur as a fan beam (see also the modeling of the 
source's pulse profiles performed by \citealt{Parmar_etal89b}).
Therefore, during the maximum of the profile (main peak) 
the angle between the column axis and the observer's line of sight has 
the highest value (the column is seen from the side).
The observer is looking almost along the beam, seeing the 
photons coming from a Compton scattering region with large optical depth. 
This leads to the observed hard power law during the peak.
Before the peak, when $E_{\rm fold}$ reaches a maximum,
the direction of the line of sight 
might be closer to that of the column axis and, thus, of the magnetic 
field lines. Due to the dependence of the scattering cross-section
on the angle between the photon direction and the magnetic field lines
\citep{HardingDaugherty91}, one expects that the photons, whose 
direction in this case is closer to that of the field lines, have 
experienced less scatterings  and, 
therefore, originated deeper inside the accretion column where the 
temperature is higher. The X-ray spectrum of these photons 
is expected to have 
larger $E_{\rm fold}$ reflecting higher electron temperature 
but a softer power-law index due to lower Compton scattering optical 
depth \citep[see e.g.][]{RybickiLightman79}, as observed.
This explanation, however, requires that in the latter case the angle
between the observer's line of sight and the column axis was not too small. 
Otherwise, depending on the geometry of the accretion flow, the column density
along the line of sight will be very high, resulting in a higher
optical depth.

The changing of the pulse profile towards the end of the outburst 
seems to confirm this picture. In the pulse profiles corresponding
to the outburst's decay
(Fig.~\ref{ppenres_too2}), one can see a new peak that appears at the phase 
interval preceeding the main pulse, i.e. where, according to our view,
the observer's line of sight is closest to the magnetic field lines.
At this pulse phase one expects to see a pencil beam if the luminosity
decreases \citep{ReigCoe98,Parmar_etal89b}.
The observed peak, therefore, may correspond to the pencil beam component
of the emission diagram, whose intensity is comaprable to that
of the fan beam component at lower luminosity.
Thus, both the pulse phase variation of the spectral continuum
during the maximum of the outburst
and the evolution of the pulse profile with luminosity fit the 
described picture where the emission diagram changes from a fan beam 
geometry close to the maximum of the outburst to a combination of a 
fan and a pencil beam at the end of the decay phase.

The behavior of the spectral parameters with pulse phase in Set~3
is less clear due to lower statistics (only four phase bins are used). 
As in the maximum of the outburst, during the main peak of the profile, 
the spectrum is characterized by a hard $\Gamma$ and a relatively 
low $E_{\rm fold}$ (Fig.~\ref{parprs_too2}). During the peak around 
phase 0.0 (pencil beam), $E_{\rm fold}$ is higher. As before, 
this can be explained by photons from the 
pencil beam moving along the magnetic field lines originating deeper in the
accretion column where the electron temperature is higher. On the other
hand, the emission region is believed to move closer to the star
surface as the luminosity decreases \citep{BaskoSunyaev76}. This means
that the column density of gas above the emission region
will be higher, leading to greater optical depth for Compton scattering. 
This can explain the hard photon index at this pulse phase.

\subsection{``Bump'' versus absorption line at 10\,keV}

As discussed in Sect.~\ref{max_prs}, to model the spectral
continuum of EXO~2030+375 in the maximum of the outburst, one has to modify 
the standard power law/cutoff model either by a ``bump'' at 
$\sim$15\,keV or by an absorption line at $\sim$10\,keV. Both models provide 
equally good fits of the pulse-phase resolved spectra. 
If interpreted as a cyclotron absorption feature \citep{Wilson_etal08}, 
the absorption line at 10\,keV is observed at rather unusual energy 
that is much smaller than the exponential cutoff energy $E_{\rm cutoff}$. 
This contrasts with all other know cyclotron line sources  
(see e.g. Fig.~9 in \citealt{Coburn_etal02}). We point out here that a 
feature around 8--15\,keV (a ``wiggle'' or a ``bump'') is observed in many 
accreting pulsars (e.g., 4U~1907, Her~X-1, \citealt{Coburn_etal02}), 
including those that do not otherwise exhibit a cyclotron line 
(e.g. GS~1843+00, \citealt{Coburn01}). Thus, the interpretation of the 
absorption line at 10\,keV in EXO~2030+375 as the cyclotron resonant 
scattering 
feature might be hasty. On the other hand, the nature of the feature 
in this and other sources is still unclear. Most probably it arises 
from modeling the spectral continuum with a simple empirical function. 
For a proper modeling of the spectrum in the considered energy range
(including the feature) one would need a proper theoretical model 
accounting for all relevant processes at the site of the X-ray emission.

\subsection{Evidence of a cyclotron line at $\sim$63\,keV}

The X-ray spectrum of EXO~2030+375 taken close to the maximum
of the outburst in the narrow pulse-phase interval (0.06--0.12) 
preceeding the main peak
shows evidence of an absorption line around $\sim$63\,keV
(Fig.~\ref{residuals}). If interpreted as a fundamental cyclotron line, 
the corresponding magnetic field strength is $B\simeq 5\times 10^{12}$\,G
($E_{\rm cyc}\sim 11.6\times (B/10^{12}\,\text{G})$\,~keV), 
which is one of the highest values among accreting pulsars. 
However, evidence of a cyclotron line at $\sim$36\,keV has previously 
been reported for EXO~2030+375 by \citet{ReigCoe99} 
during a normal outburst, so the
line at $\sim$63\,keV might well be the first harmonic rather than the
fundamental line. It is known that the relative strength of the fundamental
line and harmonics might vary significantly, sometimes making the fundamental
line more difficult to detect than the harmonic (a good example
is A~0535+26, see e.g. \citealt{Kendziorra_etal94}). 

\section{Summary and conclusions\label{concl}}

We used \textsl{INTEGRAL} observations to study the pulse-phase dependence
of the broad band X-ray spectrum of EXO~2030+375 close to the maximum 
and during the decay of its 2006 giant outburst. This is the 
first pulse-phase resolved spectral study of the source. In all observations, 
significant pulse phase variability of the X-ray continuum was observed.

Alternative spectral continuum models are discussed. We argue that the
interpretation of the feature at $\sim$10\,keV as a cyclotron absorption 
line proposed previously is questionable. 

Pulse-phase dependencies of the continuum parameters 
close to the maximum of the outburst, as well as 
the evolution of the pulse profiles from the maximum
to the end of the outburst, is qualitatively consistent
with the picture where the pulsar's emission diagram changes from the 
fan-beam configuration close to the maximum of the outburst to a 
combination of pencil and fan beams (whose amplitudes are comparable) 
at the end of the decay phase.

Evidence of an absorption line at $\sim$63\,keV is found during the 
maximum of the outburst at a narrow phase interval preceeding the main peak 
of the pulse profile. This feature can be interpreted as the first harmonic
of the previously reported cyclotron line at $\sim$36\,keV.

\begin{acknowledgements}
This research is based on observations with \textsl{INTEGRAL}, an ESA 
project with the instruments and science data center funded by ESA member 
states (especially the PI countries: Denmark, France, Germany, Italy, 
Switzerland, Spain), Czech Republic and Poland, and with the 
participation of Russia and the USA.

The work was supported by the DLR grant BA5027.

We also thank ISSI (Bern, Switzerland) for its hospitality during the 
team meetings for our collaboration.

DK thanks Valery Suleimanov (IAAT, T\"ubingen) for useful discussions.
\end{acknowledgements}

\bibliographystyle{aa}
\bibliography{refs}

\end{document}